\documentclass[aps,prl,groupedaddress,superscriptaddress, twocolumn]{revtex4-2}

\pdfoutput=1

\input{ds.def}

\newcommand{\APC}{APC, Universit\'e de Paris, CNRS, Astroparticule et Cosmologie, Paris F-75013, France}
\newcommand{\AQLNGS}{INFN Laboratori Nazionali del Gran Sasso, Assergi (AQ) 67100, Italy}
\newcommand{\AQGSSI}{Gran Sasso Science Institute, L'Aquila 67100, Italy}

\newcommand{\AstroCeNT}{AstroCeNT, Nicolaus Copernicus Astronomical Center, 00-614 Warsaw, Poland}
\newcommand{\Augustana}{Physics Department, Augustana University, Sioux Falls, SD 57197, USA}
\newcommand{\Belgorod}{Radiation Physics Laboratory, Belgorod National Research University, Belgorod 308007, Russia}
\newcommand{\BHSU}{School of Natural Sciences, Black Hills State University, Spearfish, SD 57799, USA}

\newcommand{\CAUniPHY}{Physics Department, Universit\`a degli Studi di Cagliari, Cagliari 09042, Italy}
\newcommand{\CAINFN}{INFN Cagliari, Cagliari 09042, Italy}

\newcommand{\CPPM}{Centre de Physique des Particules de Marseille, Aix Marseille Univ, CNRS/IN2P3, CPPM, Marseille, France}
\newcommand{\CTLNS}{INFN Laboratori Nazionali del Sud, Catania 95123, Italy}
\newcommand{\ENUniCEE}{Engineering and Architecture Faculty, Universit\`a di Enna Kore, Enna 94100, Italy}

\newcommand{\FNAL}{Fermi National Accelerator Laboratory, Batavia, IL 60510, USA}

\newcommand{\GEUni}{Physics Department, Universit\`a degli Studi di Genova, Genova 16146, Italy}
\newcommand{\GEINFN}{INFN Genova, Genova 16146, Italy}

\newcommand{\Hawaii}{Department of Physics and Astronomy, University of Hawai'i, Honolulu, HI 96822, USA}
\newcommand{\Houston}{Department of Physics, University of Houston, Houston, TX 77204, USA}
\newcommand{\IHEP}{Institute of High Energy Physics, Beijing 100049, China}

\newcommand{\JINR}{Joint Institute for Nuclear Research, Dubna 141980, Russia}
\newcommand{\Krakow}{M. Smoluchowski Institute of Physics, Jagiellonian University, 30-348 Krakow, Poland}
\newcommand{\Kurchatov}{National Research Centre Kurchatov Institute, Moscow 123182, Russia}

\newcommand{\LNFINFN}{INFN Laboratori Nazionali di Frascati, Frascati 00044, Italy}
\newcommand{\LNHB}{Universit\'e Paris-Saclay, CEA, List, Laboratoire National Henri Becquerel (LNE-LNHB), F-91120 Palaiseau, France}

\newcommand{\LPNHE}{LPNHE, CNRS/IN2P3, Sorbonne Universit\'e, Universit\'e Paris Diderot, Paris 75252, France}
\newcommand{\Manchester}{The University of Manchester, Manchester M13 9PL, United Kingdom}
\newcommand{\MEPhI}{National Research Nuclear University MEPhI, Moscow 115409, Russia}

\newcommand{\MIINFN}{INFN Milano, Milano 20133, Italy}

\newcommand{\MIUni}{Physics Department, Universit\`a degli Studi di Milano, Milano 20133, Italy}
\newcommand{\MSU}{Skobeltsyn Institute of Nuclear Physics, Lomonosov Moscow State University, Moscow 119234, Russia}
\newcommand{\NAINFN}{INFN Napoli, Napoli 80126, Italy}
\newcommand{\NAUniPHY}{Physics Department, Universit\`a degli Studi ``Federico II'' di Napoli, Napoli 80126, Italy}

\newcommand{\Petersburg}{Saint Petersburg Nuclear Physics Institute, Gatchina 188350, Russia}
\newcommand{\PGUniCBB}{Chemistry, Biology and Biotechnology Department, Universit\`a degli Studi di Perugia, Perugia 06123, Italy}
\newcommand{\PGINFN}{INFN Perugia, Perugia 06123, Italy}
\newcommand{\PIINFN}{INFN Pisa, Pisa 56127, Italy}
\newcommand{\PIUniPHY}{Physics Department, Universit\`a degli Studi di Pisa, Pisa 56127, Italy}
\newcommand{\PNNL}{Pacific Northwest National Laboratory, Richland, WA 99352, USA}
\newcommand{\Princeton}{Physics Department, Princeton University, Princeton, NJ 08544, USA}

\newcommand{\RHUL}{Department of Physics, Royal Holloway University of London, Egham TW20 0EX, UK}
\newcommand{\RMTreINFN}{INFN Roma Tre, Roma 00146, Italy}
\newcommand{\RMTreUni}{Mathematics and Physics Department, Universit\`a degli Studi Roma Tre, Roma 00146, Italy}
\newcommand{\RMUnoINFN}{INFN Sezione di Roma, Roma 00185, Italy}
\newcommand{\RMUnoUni}{Physics Department, Sapienza Universit\`a di Roma, Roma 00185, Italy}

\newcommand{\SSUniCHP}{Chemistry and Pharmacy Department, Universit\`a degli Studi di Sassari, Sassari 07100, Italy}

\newcommand{\UCDavis}{Department of Physics, University of California, Davis, CA 95616, USA}
\newcommand{\UCLA}{Physics and Astronomy Department, University of California, Los Angeles, CA 90095, USA}
\newcommand{\UMass}{Amherst Center for Fundamental Interactions and Physics Department, University of Massachusetts, Amherst, MA 01003, USA}

\newcommand{\USP}{Instituto de F\'isica, Universidade de S\~ao Paulo, S\~ao Paulo 05508-090, Brazil}
\newcommand{\VTech}{Virginia Tech, Blacksburg, VA 24061, USA}
\newcommand{\kings}{Physics, Kings College London, Strand, London WC2R 2LS, UK}

\begin{document}
\title{Search for dark matter-nucleon interactions via Migdal effect with \DSf}

\author{P.~Agnes}\affiliation{\RHUL}
\author{I.F.M.~Albuquerque}\affiliation{\USP}
\author{T.~Alexander}\affiliation{\PNNL}
\author{A.K.~Alton}\affiliation{\Augustana}
\author{M.~Ave}\affiliation{\USP}
\author{H.O.~Back}\affiliation{\PNNL}
\author{G.~Batignani}\affiliation{\PIINFN}\affiliation{\PIUniPHY}
\author{K.~Biery}\affiliation{\FNAL}
\author{V.~Bocci}\affiliation{\RMUnoINFN}
\author{W.M.~Bonivento}\affiliation{\CAINFN}
% ...

\collaboration{The DarkSide Collaboration}\noaffiliation


\begin{abstract}
Dark matter elastic scattering off nuclei can result in the excitation and ionization of the recoiling atom through the so-called Migdal effect. The energy deposition from the ionization electron adds to the energy deposited by the recoiling nuclear system and  allows for the detection of interactions of sub-GeV/c$^2$ mass dark matter. 
We present new constraints for sub-GeV/c$^2$ dark matter using the dual-phase liquid argon time projection chamber of the  DarkSide-50 experiment with an exposure of \DSfDdExposureNew.
The analysis is based on the ionization signal alone and significantly enhances the sensitivity of \DSf, enabling sensitivity to dark matter with masses down to~40~MeV/c$^2$. Furthermore, it sets the most stringent upper limit on the spin independent dark matter nucleon cross section for masses below $3.6$~GeV/c$^2$.
\end{abstract}

\maketitle

The presence of dark matter (DM) in the universe is strongly supported by many observations \cite{Rubin:1980zd,Clowe:2003tk,Planck:2018vyg}, based only on DM gravitational effects.
Other possible interactions remain unknown. Weakly interacting massive particles are theoretically-favored DM candidates with masses in the GeV/c$^2$--TeV/c$^2$ range~\cite{Jungman:1995df}. Attempts to detect DM elastic scattering off target nuclei have resulted in strong limits on DM interactions 
for masses above a few GeV/c$^2$ \cite{DarkSide:2015cqb,LUX:2016ggv,PandaX-II:2017hlx,XENON:2018voc}. 
Furthermore, several mechanisms that explain the observed DM density point to light DM particles (LDM), with masses in the sub-GeV/c$^2$ range~\cite{Essig:2013lka,Battaglieri:2017aum}. 

LDM is difficult to probe with direct detection experiments because the DM-induced nuclear recoil (NR) energy is generally below the detection threshold.
However, atomic effects modeled by Migdal~\cite{Migdal:1941} predict emission of electrons associated with a fraction of nuclear recoils. This electron recoil (ER) component, in addition to the NR one, increases the probability of exceeding the detection threshold, thus opening a window of exploration for DM particles with masses down to a few tens of MeV/c$^2$. 
The idea by Migdal originated in the context of nuclear physics for alpha and beta emissions \cite{RUIJGROK1983537,Vegh_1983,Baur_1983,Sharma:2017fmo}, and has been recently adapted to direct dark matter experiments \cite{Bernabei:2007jz,Ibe:2017yqa,Dolan:2017xbu,Bell:2019egg,Baxter:2019pnz,Essig:2019xkx,Liang:2019nnx,GrillidiCortona:2020owp,Liu:2020pat,Dey:2020sai,Knapen:2020aky,Bell:2021zkr,Acevedo:2021kly,Wang:2021oha}. 

In this Letter, we report the results of a search for LDM-nucleon elastic interactions based on the ionization signal in the \DSf\ (\DSfs) detector, taking into account the extra energy detected due to the Migdal effect (ME). Previous DM searches including the ME were performed by several Collaborations~\cite{LUX:2018akb, EDELWEISS:2019vjv,CDEX:2019hzn, XENON:2019zpr,COSINE-100:2021poy}.

\DSfs\ uses a dual-phase liquid argon (LAr) time projection chamber (TPC), located at the INFN Laboratori Nazionali del Gran Sasso (LNGS) in Italy. Particle interactions in the 46.4$\pm$0.7 kg active target induce scintillation pulses (\SOne) and ionization electrons. The latter are drifted through an electric field up to the gas pocket, at the top of the TPC, where they produce a secondary pulse of light (\STwo) by electroluminescence. 
\SOne\ and \STwo\ ultraviolet photons are converted into the visible range by tetraphenyl butadiene, a wavelength shifter that coats the inner surfaces of the TPC. Visible photons are 
detected by two arrays of 19 3-in photomultipliers, one located above the anode and one below the cathode, respectively. The TPC is installed at the center of a stainless-steel sphere, filled with 30 t of boron-loaded liquid scintillator, which is in turn installed in a cylindrical tank, filled with 1 kt of ultra-pure water. The scintillator and water detectors are equipped with PMTs, and act as neutron and muon veto, respectively. More details on the detector can be found in ref.~\cite{DarkSide:2014llq,DarkSide:2015cqb,DarkSide:2018bpj,DarkSide:2018kuk,DarkSide:2018ppu}.

We perform this analysis using the \DSfDdLTPostQualCutnumNew\ live-days \DSfs\ dataset, from December 12, 2015 to October 4, 2017.
We use the ionization signal \STwo\ since it has significantly lower detection threshold than \SOne\ thanks to the gain of the electroluminescence.
The region of interest (ROI) is defined as where the ionization response is calibrated~\cite{DarkSide:2021bnz} and backgrounds are well-understood. We characterize the strength of the ionization signal by the number of electrons that are extracted into the gas-region at the top of the TPC. Given the electric field settings of the TPC, the extraction efficiency for electrons from the liquid into the gas is essentially 100\% and so $N_{e}$ is a good measure of the ionization signal. This corresponds to the number $N_e$ of electrons counted in \STwo\ within [4,\,170]~e$^-$, equivalent to [0.06,\,21] keV$_{\textrm{er}}$ ([0.64,393] keV$_{\textrm{nr}}$) in the ER (NR) energy scale. Above $4e^-$, the contribution of spurious electrons, captured by impurities along their drift and re-emitted with a delay, to the background model is negligible~\cite{long_paper}. 

We consider only single-scatter events occurring in the central fiducial mass of 19.4$\pm$0.3 kg. Such events are identified by requiring a single valid \STwo\ pulse. The extra \STwo\ pulses induced by electrons extracted from the cathode by the UV photons from \SOne\ or \STwo\ pulses are identified as echoes by their timing and not counted.
A set of quality cuts, based on the topological distribution and the time profile of the \STwo\ signal, and on \STwo/\SOne, is implemented to reject events with overlapping pulses without appreciable loss of acceptance, as described in Ref.~\cite{long_paper}.
An additional set of selection cuts is applied to remove spurious \STwo\ pulses mainly induced by electrons captured by impurities, events with an echo from surface alphas that lose normal \STwo\ to the TPC wall, and pile up events, 
associated to random coincidences between very low \SOne\ and \STwo\ pulses from the anode.
The final data-set accounting for the quality and selection cuts corresponds to an exposure of \DSfDdExposureNew.  The overall acceptance,   almost flat with respect to the recoil energy,  varies from 38\% at 4~e$^-$ to 40\% at higher than  15~e$^-$. 

The major sources of background events in the ROI  and in the fiducial volume are $^{39}$Ar and $^{85}$Kr  decays occurring in the LAr bulk, whose rates are expected 
to be (6.5$\pm$0.9)$\times$10$^{-4}$~Hz and (1.7$\pm$0.1)$\times$10$^{-3}$~Hz, respectively, and  $\gamma$-rays and X-rays from radioactive contaminants in the PMTs   and stainless-steel cryostat, which contribute at  (3.5$\pm$0.4)$\times$10$^{-3}$~Hz and  (5.9$\pm$0.4)$\times$10$^{-4}$~Hz, respectively \cite{long_paper}.
Backgrounds originating from radiogenic and cosmogenic neutrons, as well as coherent elastic neutrino-nucleus scattering from solar and atmospheric neutrinos, are negligible in comparison.
The main systematic uncertainties for the    $^{39}$Ar and $^{85}$Kr background stem from the atomic exchange, screening effects, and ionization response. A subdominant systematic uncertainty from the Q-value is also included~\cite{Wang:2021xhn}. The systematic uncertainties and their impact are discussed in detail in~\cite{long_paper}. 
Uncertainties on the PMT and cryostat  backgrounds are due to the detector response and from Monte Carlo statistics. More details on the event selection and background models are described in~\cite{long_paper}.

The calibration of the detector and its response to ER and NR energy deposits has been performed in~\cite{DarkSide:2021bnz}. The ionization response to electronic recoils has been measured down to 180 eV$_{er}$ and a fit to the data with
a function of the Thomas-Imel box model form 
allows an extrapolation down to $\mathcal{O}(100\:{\rm eV})$.
Similarly, the expected number of ionization electrons for NR is given by the Thomas-Imel box model, where the number of electron-ion pairs is obtained with Bezrukov's model \cite{Bezrukov:2010qa} and with the Ziegler et al. model for the nuclear screening function~\cite{ziegler:1982}. The ionization response to NR has been measured down to 500 eV$_{nr}$. This is the lowest threshold ever reached in liquid argon and corresponds to 3 ionization electrons. The ER and NR ionization models are constrained by fitting the $^{241}$Am$^9$Be and $^{241}$Am$^{13}$C neutron sources data, $\beta$-decay data of $^{39}$Ar, and electron captures of $^{37}$Ar obtained during the \DSfs\ calibration campaign, and by external datasets from the SCENE \cite{SCENE:2014iyj}, ARIS \cite{Agnes:2018mvl} and Joshi \emph{et al.} \cite{Joshi:2014fna} experiments. Details can be found in \cite{DarkSide:2021bnz}.

The elastic scattering of a DM particle off an argon nucleus at rest induces an instantaneous momentum change of the nucleus with respect to the atomic electrons, resulting in the possible ionization or excitation of the atom: this is the ME.
When considering the ME, both NR and ER signals are present. For the first time, we consider and sum both contributions to the predicted signal.

The differential event rate for DM elastically scattering on an argon nucleus with respect to the nuclear recoil energy $E_{nr}$ and DM velocity $v$ is given by
\begin{equation}
\frac{d^2R_{\textrm{nr}}}{dE_{\textrm{nr}}\, dv} = \frac{\rho_{DM}\,\sigma_{SI}}{2\, \mu_N^2 \,m_{DM}}\frac{f(v)}{v},
\end{equation}
where $\rho_{DM}=0.3$ GeV cm$^{-3}$ c$^{-2}$ is the local DM density, $m_{DM}$ is its mass, $\sigma_{SI}$ is the DM-nucleus spin independent scattering cross section, $\mu_N$ is the DM-nucleus reduced mass, and $f(v)$ is the DM speed distribution in the laboratory frame. 
We assume the Standard Halo Model with a DM escape velocity $v_{esc}=544$ km/s, and local standard of rest velocity $v_{0}=238$ km/s~\cite{ShawnWhitePaper}.

The rate for a nuclear recoil energy $E_{\textrm{nr}}$, accompanied by an ionization electron with energy $E_{er}$ is given by \cite{Ibe:2017yqa} 
\begin{equation}
\frac{d^3R}{dE_{\textrm{nr}}\,dE_{\textrm{er}}\,dv} = \frac{1}{2\pi}\sum_{n,\ell} \frac{d^2R_{\textrm{nr}}}{dE_{\textrm{nr}}\,dv}\frac{dp^c_{q_e}(n\ell\to E_{\textrm{er}})}{dE_{\textrm{er}}},
\label{eq:d2R_Migdal}
\end{equation}
where $dR_{\textrm{nr}}/dE_{\textrm{nr}}$ is the standard DM nuclear recoil rate, $p_{q_e}^c$ is the probability to emit an electron from the ($n, \ell$) shell with final energy $E_{\textrm{er}}$, $q_e=m_e \sqrt{2 E_{nr}/m_N}$ is the electron momentum in the nucleus rest frame immediately after the DM interaction, $m_e$ is the electron mass, and $m_N$ is the nucleus mass.
Since the emitted electron may come from an inner orbital, the remaining excited state will immediately release further energy in the form of additional electrons or photons. These are measured simultaneously with the energy deposited by the initial ionization electron. As a consequence, the total energy deposited in the electromagnetic channel can be estimated to be $E_{EM}=E_{\textrm{er}}+E_{n\ell}$, where $E_{n\ell}$ is the binding energy of the $(n,\,\ell)$ state.  In this analysis, we use the differential probabilities for isolated Ar atoms computed in \cite{Ibe:2017yqa}, and we consider the ionization contributions of all the electron shells.   The fraction of events where the ME occurs increases as the DM mass increases. As an example, these fractions are $2.9\times10^{-5}$ at 100~MeV/c$^2$ and $1.2\times10^{-3}$ at 1~GeV/c$^2$.

In Ref. \cite{Liu:2020pat} it has been shown that the prediction of Ref. \cite{Ibe:2017yqa} for the probability of emitting an electron from the 
valence shell in isolated argon atoms is robust.
However, in liquid argon, the valence shell shows a band structure and a reduced binding energy.
Neglecting this difference in the computation results in a smaller ionization probability~\cite{Catena:2019gfa}, thus reducing the predicted ME signal event yield.

\begin{figure}[t]
 \includegraphics[width=\columnwidth]{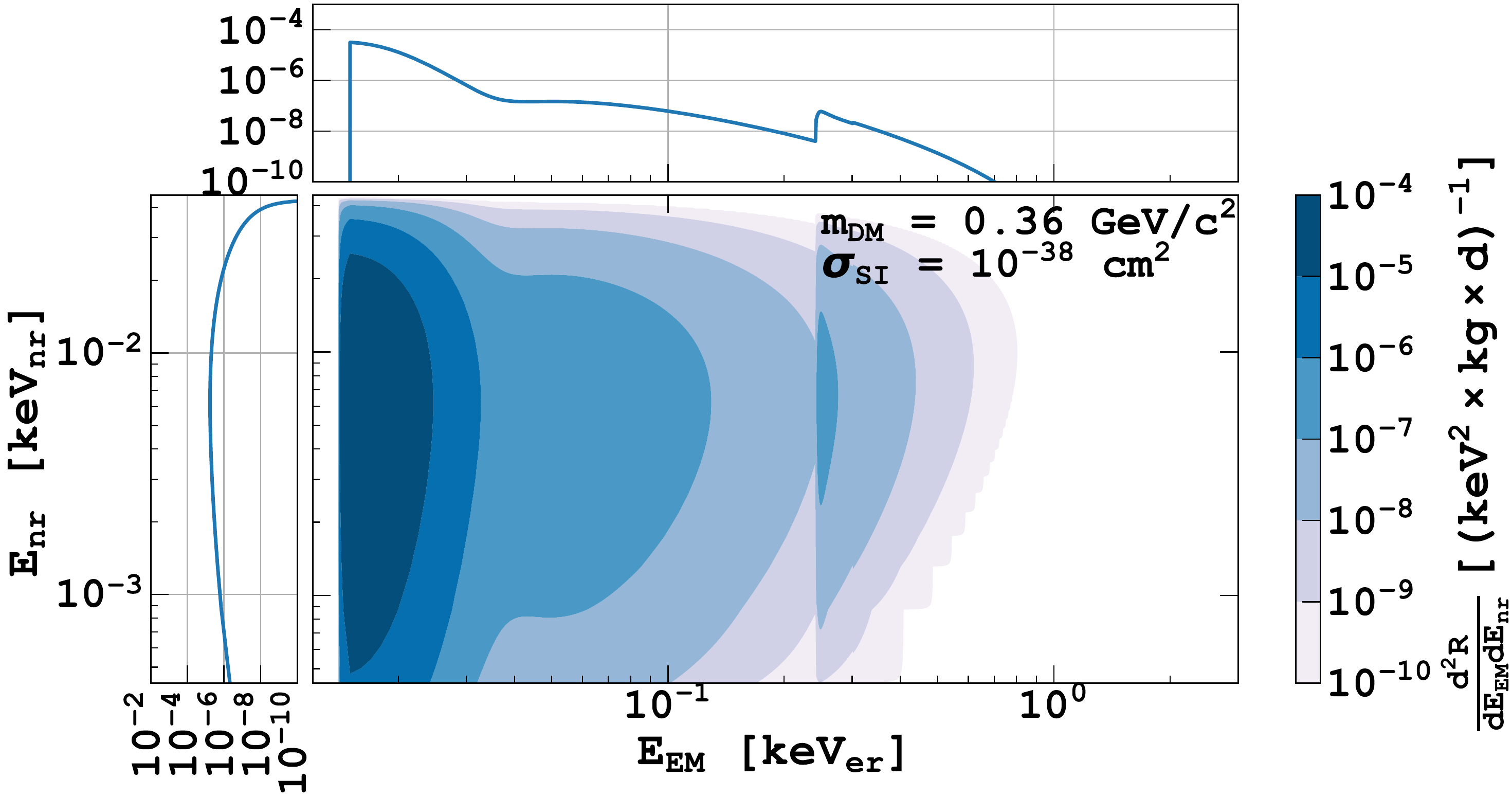}
\includegraphics[width=\columnwidth]{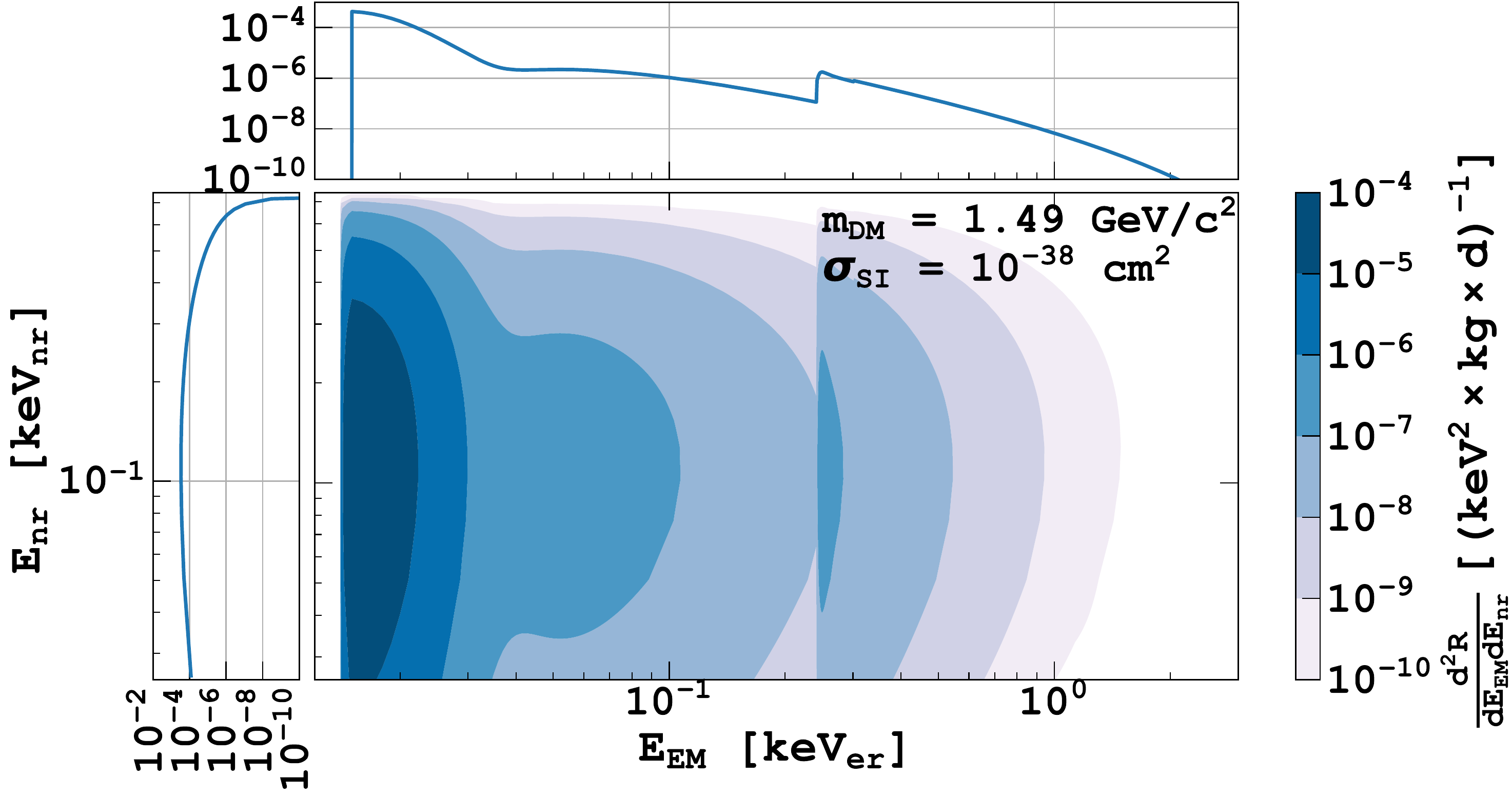}
 \caption{
 2D differential ionization rate as a function of the nuclear recoil energy ($E_{\textrm{nr}}$) and electromagnetic channel energy ($E_{\textrm{EM}}$) is shown for two representative DM masses, 0.36~GeV/c$^2$ (top) and 1.49~GeV/c$^2$ (bottom).
    The rate is given in events/(keV$^{2}$ kg d) and covers $E_{EM}$ from 0.01 to 3 keV and $E_{nr}$ from $4\cdot 10^{-4}$  to $4\cdot 10^{-2}$ keV for a DM mass of $0.36$ GeV/c$^{2}$ and $E_{nr}$ from $2.5\cdot 10^{-2}$ to $7.5\cdot 10^{-1}$ keV for a particle of mass 1.49 GeV/$c^{2}$.
    The top and side panels of each figure depict the corresponding 
    integrated distributions in the ME electron (top panels) and the NR (left panels) channels.
    }
\label{fig:maps}
\end{figure} 
The 2D  differential rate in Eq.~\eqref{eq:d2R_Migdal} for two representative DM masses (0.36~GeV/c$^2$ and 1.49~GeV/c$^2$)
as a function of $E_{EM}$ and $E_{\textrm{nr}}$
are shown in Figure~\ref{fig:maps}, along with the corresponding 1D integrated distributions in the ME electron and the NR channels. The peaks in the ME electron spectrum correspond to the contribution of the different atomic shells, with binding energies $E_{n\ell}$ from Ref.~\cite{Ibe:2017yqa}.

The signal for spin independent DM-nucleon scattering is modeled with a Monte Carlo approach simulating the event resulting from the combination  of the recoiling atom and  the ionization electron induced by the ME. The detector response model is applied independently to the corresponding NR and  ER components, accounting for the ionization and electron-ion recombination processes \cite{DarkSide:2021bnz}. The ER component is modelled as a single energy deposit, despite being the sum of primary ionization ($E_{\textrm{er}}$) and subsequent X-ray/Auger cascade ($E_{n\ell}$). We tested this assumption against an alternative description of the ME process, assumed as the results of two independent energy deposits of $E_{\textrm{er}}$ and $E_{n\ell}$, and find that the calculated exclusion limits are indistinguishable.

Regarding NRs, they  are subject to the quenching effect, a stochastic process whose statistics governing its fluctuation  is unknown.  For this reason, we considered two models where quenching fluctuations are either suppressed (NQ) or binomial (QF).

\begin{figure}[t]
 \includegraphics[width=0.5\textwidth]{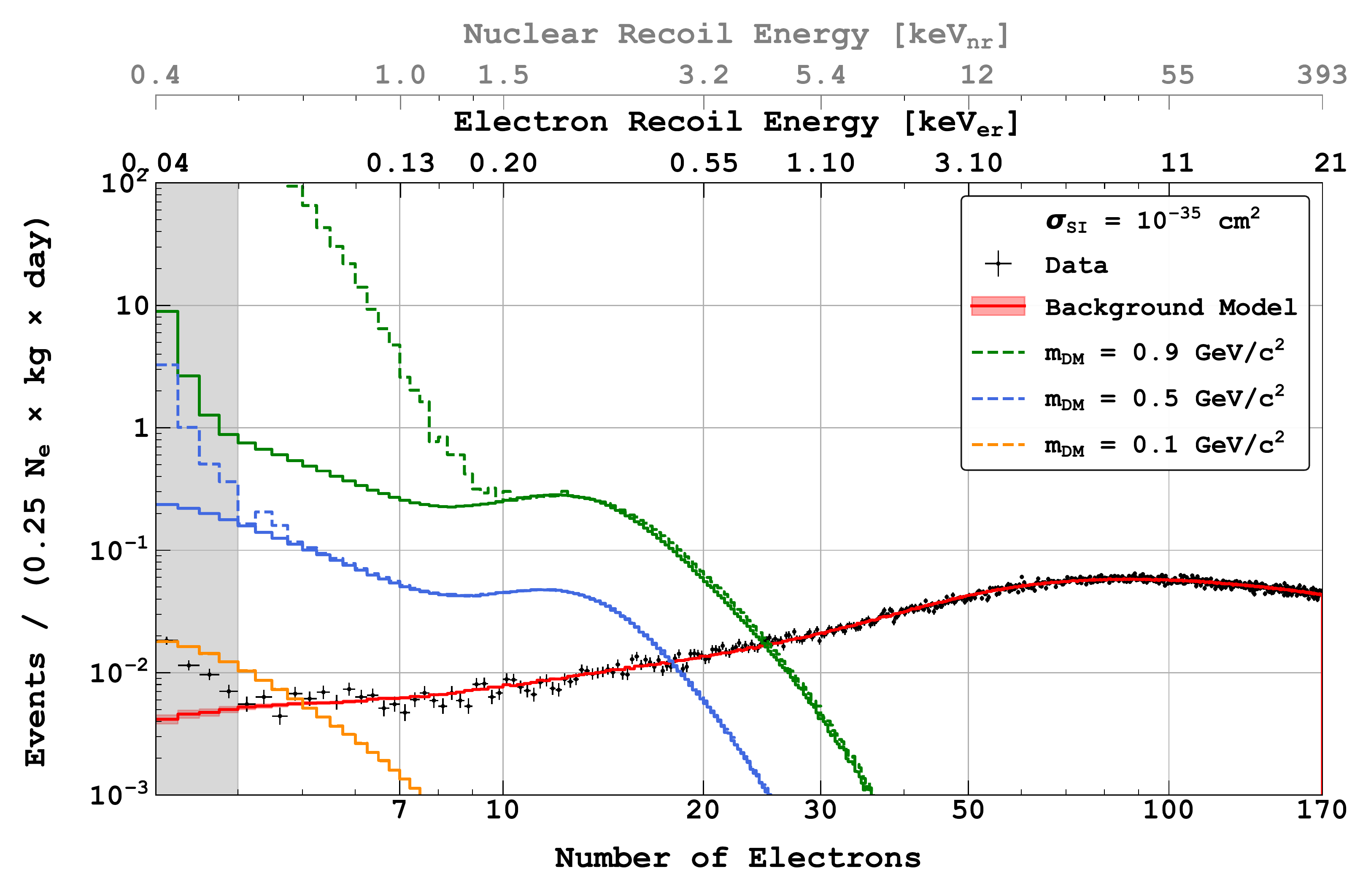}
 \caption{Data (black) and background model (red) after the selection and fit described in \cite{long_paper}. The expected spectra 
 including the Migdal effect assuming a spin independent DM-nucleon scattering cross section of $10^{-35}$ cm$^2$ and DM masses of 0.1, 0.5 and 0.9 GeV/c$^2$ are shown in orange, blue and green. 
 The gray shaded band shows the \STwo\ threshold used in the analysis.}
\label{fig:signal_rate}
\end{figure}

The predicted dark matter signals (orange, blue and green lines) for both QF (dashed) and NQ (solid) models are shown in Figure \ref{fig:signal_rate} together with   the \DSfs\ data (black points) and the fitted  background model (red curve). The signal shown was produced for a spin independent scattering cross section $\sigma_{DM}=10^{-35}$ cm$^2$ and different dark matter masses (orange for $m_{DM} = 0.1$ GeV/c$^2$, blue for $m_{DM} = 0.5$ GeV/c$^2$ and green for $m_{DM} = 0.9$ GeV/c$^2$).

The signal rate for $m_{DM}=0.9$ GeV/c$^2$ contains contributions from NR and ME which are both above the analysis threshold when the quenching fluctuations are included, with the two contributions  combined in order to set the limit. On the other hand, the 
distributions for $m_{DM}=0.5$ and $0.1$ GeV/c$^2$ are dominated above threshold solely by the ME, independently of the fluctuation model chosen.

The \STwo\ observed energy spectrum is interpreted using a binned profile likelihood
as described in detail in~\cite{long_paper}. The bins are assumed independent of each other and in each bin the probability is described by a Poisson distribution. The Poisson intensity parameter of the $i$-th bin is given by the sum of the signal contribution,
multiplied by its normalization parameter,
and the expected background templates. These quantities are affected by the uncertainties on the exposure, ionization energy scale, 
the estimate of the radioactivity present in the detector, and  the calculations of atomic exchange and screening effects impacting  $^{39}$Ar and $^{85}$Kr  first forbidden unique beta decay spectral shapes. Such systematic effects are implemented by means of a set of nuisance parameters that acts on the normalizations and spectral shapes of the background and signal spectra. 
This likelihood has been used to perform a background-only fit in the region $N_{e}=[4, 170]$, resulting in a good description of the observed spectrum as shown by the red histogram of Figure \ref{fig:signal_rate}. The post-fit values of the nuisance parameters are in good agreement with the nominal ones \cite{long_paper}, confirming the reliability of the fit.

The search for spin independent dark matter-nucleon interactions via the ME is performed with a profile log-likelihood ratio test statistic based on the above likelihood function and the dark matter signal described in the previous paragraphs.

\begin{figure}[t]
 \includegraphics[width=0.48\textwidth]{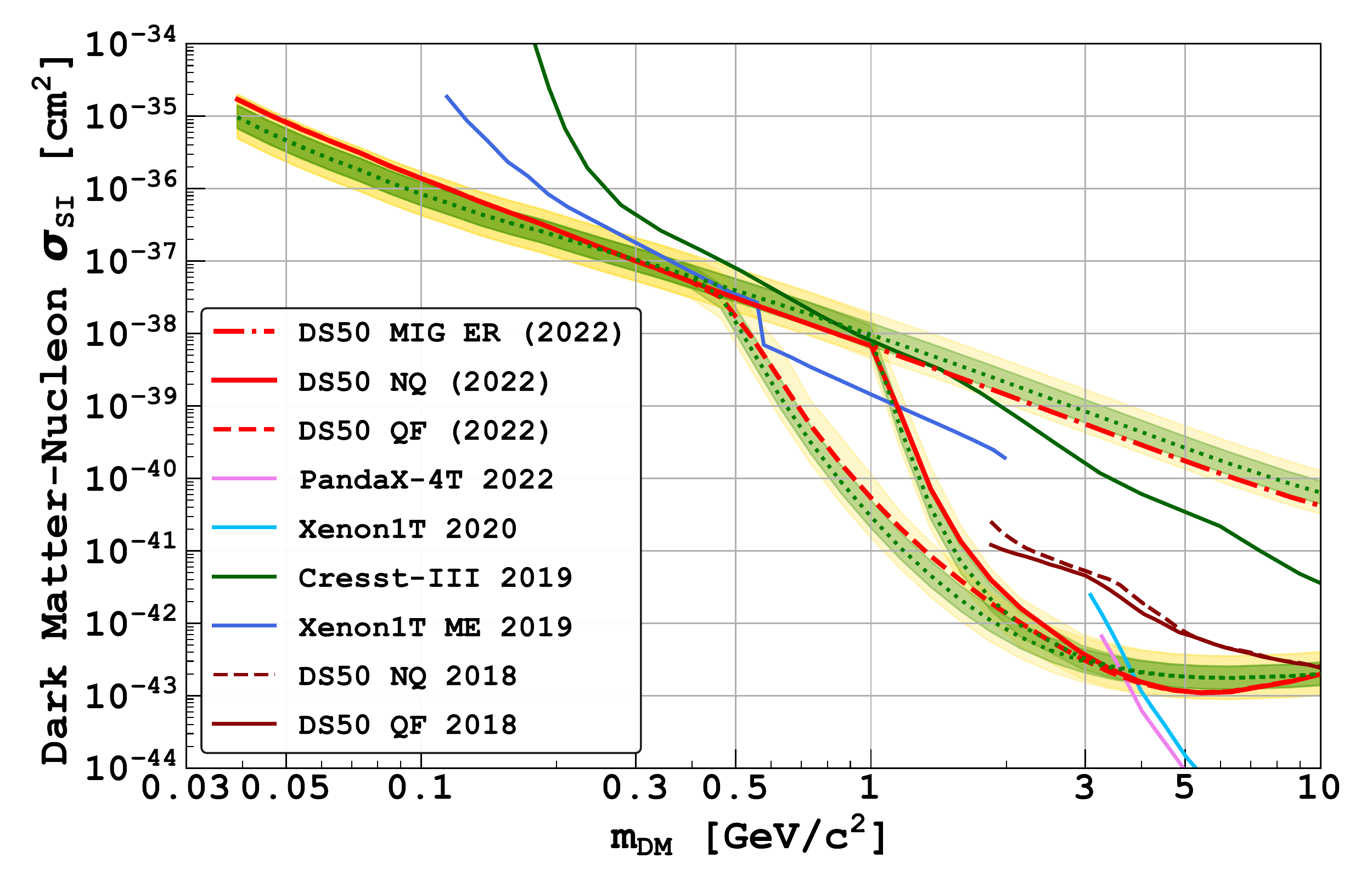}
 \caption{Upper limits on the spin independent DM-nucleon cross section at 90$\%$ C.L. obtained with a signal including the Migdal effect, together with the corresponding $\pm1\sigma$ (green shaded area) and $\pm2\sigma$ (yellow shaded area) expected limits. NQ is red solid, QF is dashed, and the ER contribution from the ME is dash-dotted.
 Also shown are limits 
 Cresst-III (green) \cite{CRESST:2019jnq}, 
 Xenon1T (light and dark blue) \cite{XENON:2019gfn,XENON:2019zpr},  PandaX-4T \cite{PandaX-4T:2021bab} and \DSfs\ (dark red) \cite{DarkSide:2018bpj}. Other weaker limits \cite{SuperCDMS:2013eoh,CRESST:2015txj,LUX:2016ggv,Behnke:2016lsk,PICO:2017tgi,SuperCDMS:2017nns,XENON:2018voc,PandaX-4T:2021bab,DAMIC:2021esz} and claimed discovery \cite{CoGeNT:2010ols, Angloher:2011uu, CDMS:2013juh, Bernabei:2013xsa} are not shown.}
\label{fig:limits_obs_exp}
\end{figure}

The observed limit at $90\%$ C.L. for the NQ (QF) signal model is shown as a solid (dashed) red curve in Figure \ref{fig:limits_obs_exp}, together with the corresponding $\pm1\sigma$ (green shaded area) and $\pm2\sigma$ (yellow shaded area) expected limits. The observed limit is compatible within $1\sigma$ with the expected one, showing no significant excess above the expected background in the region above $N_e=4$.
The choice of the fluctuation model affects only the intermediate region between 0.5 and 5 GeV/c$^2$. Indeed, this is the transition region between a signal that is dominated by the nuclear recoil and one that is dominated by the ME with nuclear recoils just below the analysis threshold. 
The overlap between ER and NR ion-electron clouds, if spatially close, may reduce the number of free electrons. Such an effect is not accounted for in this work. However, the maximal size of this effect can be inferred by comparing the obtained limit with the one evaluated by assuming  the ME-induced ER component only (dot-dashed).

The observed upper limits presented in this Letter are compared in Figure \ref{fig:limits_obs_exp} with other experiments \cite{XENON:2019gfn,XENON:2019zpr,PandaX-4T:2021bab,CRESST:2019jnq,DarkSide:2018bpj}.  
The limit is entirely driven by the ME for DM masses below $0.5\:{\rm GeV/c}^2$.
The \DSfs\ experiment reaches the best sensitivity for the dark matter spin-independent scattering cross section for masses below 3.6~GeV/c$^2$,  
improving considerably the sensitivity with respect to the analysis of 2018~\cite{DarkSide:2018bpj}.

The limits benefit from the extended signal region $N_e \in [4,\,170]$ even though the signal rate typically is negligible with respect to the background rate for $N_e>50$ for masses of $\mathcal{O}(1)$~GeV. The larger $N_e$ region provides further constraints on the calibration parameters  and background model. 

Exploiting data from the full exposure of the \DSfs\ experiment, we performed a search for LDM by analysing the ionization signals induced by DM particles scattering off nuclei, enhanced by the Migdal effect. The Migdal detection channel, together with the new calibration \cite{DarkSide:2021bnz}, data selection, and background model \cite{long_paper}, improves significantly the sensitivity of \DSfs\ to spin-independent DM-nucleon interactions for sub-GeV masses. 
This analysis sets the world best limit on the spin-independent DM-nucleon cross section for masses below $3.6$ GeV/c$^2$ and down to 40 MeV/c$^2$. The same analysis approach  was also applied  to  improve existing  limits on dark matter particle interactions with electron final states \cite{DarkSide-50:2022hin}. With the DarkSide-20k detector under construction at the LNGS~\cite{DarkSide-20k:2017zyg}, we hope to improve on all these limits significantly.

\begin{acknowledgements}
The DarkSide Collaboration offers its profound gratitude to the LNGS and its staff for their invaluable technical and logistical support. We also thank the Fermilab Particle Physics, Scientific, and Core Computing Divisions. Construction and operation of the DarkSide-50 detector was supported by the U.S. National Science Foundation (NSF) (Grants No. PHY-0919363, No. PHY-1004072, No. PHY-1004054, No. PHY-1242585, No. PHY-1314483, No. PHY-1314501, No. PHY-1314507, No. PHY-1352795, No. PHY-1622415, and associated collaborative grants No. PHY-1211308 and No. PHY-1455351), the Italian Istituto Nazionale di Fisica Nucleare, the U.S. Department of Energy (Contracts No. DE-FG02-91ER40671, No. DEAC02-07CH11359, and No. DE-AC05-76RL01830), the Polish NCN (Grant No. UMO-2019/33/B/ST2/02884) and the Polish Ministry for Education and Science (Grant No. 6811/IA/SP/2018). We also acknowledge financial support from the French Institut National de Physique Nucl\'eaire et de Physique des Particules (IN2P3),   the  IN2P3-COPIN consortium (Grant No. 20-152),  and the UnivEarthS LabEx program (Grants No. ANR-10-LABX-0023 and No. ANR-18-IDEX-0001),  from the São Paulo Research Foundation (FAPESP) (Grant No. 2016/09084-0),  from the Interdisciplinary Scientific and Educational School of Moscow University ``Fundamental and Applied Space Research'',  from the Program of the Ministry of Education and Science of the  Russian  Federation  for  higher  education  establishments,  project No. FZWG-2020-0032 (2019-1569),  from IRAP AstroCeNT funded by FNP from ERDF, and from the Science and Technology Facilities Council, United Kingdom.  I.~Albuquerque is partially supported by the Brazilian Research Council (CNPq). This project has received funding from the European Union’s Horizon 2020 research and innovation program under grant agreement No 952480. Isotopes used in this research were supplied by the United States Department of Energy Office of Science by the Isotope Program in the Office of Nuclear Physics.
 \end{acknowledgements}

\bibliography{biblio}
\end{document}